\documentclass[aps,prl,twocolumn,superscriptaddress,showpacs,floatfix]{revtex4-1}

\usepackage{graphicx}

\newcommand{\al}{$\alpha$}
\newcommand{\g}{$\gamma$}

\newcommand{\rag}{($\alpha$,$\gamma$)}
\newcommand{\ran}{($\alpha$,n)}

\newcommand{\rann}{($\alpha$,2n)}

\newcommand{\rgn}{($\gamma$,n)}

\newcommand{\rga}{($\gamma$,$\alpha$)}
\newcommand{\rgp}{($\gamma$,p)}

\newcommand{\stot}{$\sigma_{\rm{reac}}$}

\newcommand{\Nsv}{$N_A$$\left< \sigma v \right>$}

\newcommand{\rpro}{$r$-process}

\newcommand{\gpro}{$\gamma$-process}

\newcommand{\pnuc}{$p$-nucleus}
\newcommand{\pnucs}{$p$-nuclei}
\newcommand{\sfact}{S-factor}

\newcommand{\auvii}{$^{197}$Au}

\newcommand{\wvi}{$^{176}$W}
\newcommand{\wnull}{$^{180}$W}
\newcommand{\osnull}{$^{180}$Os}
\newcommand{\niviii}{$^{58}$Ni}
\newcommand{\zniv}{$^{64}$Zn}

\begin{document}

\title{Successful prediction of total $\alpha$-induced reaction cross sections
  at astrophysically relevant sub-Coulomb energies using a novel approach
}

\author{P.~Mohr}%
\email{mohr@atomki.mta.hu}%
\affiliation{Institute for Nuclear Research (MTA Atomki), H--4001 Debrecen, Hungary}
\affiliation{Diakonie-Klinikum, D--74523 Schw\"abisch Hall, Germany}

\author{Zs.~F\"ul\"op}%
\affiliation{Institute for Nuclear Research (MTA Atomki), H--4001 Debrecen, Hungary}

\author{Gy.~Gy\"urky}%
\affiliation{Institute for Nuclear Research (MTA Atomki), H--4001 Debrecen, Hungary}

\author{G.~G.~Kiss}%
\affiliation{Institute for Nuclear Research (MTA Atomki), H--4001 Debrecen, Hungary}

\author{T.~Sz\"ucs}%
\affiliation{Institute for Nuclear Research (MTA Atomki), H--4001 Debrecen, Hungary}%

\date{\today}

\begin{abstract}
The prediction of stellar ($\gamma$,$\alpha$) reaction rates for heavy nuclei
is based on the calculation of ($\alpha$,$\gamma$) cross sections at
sub-Coulomb energies. These rates are essential for modeling the
nucleosynthesis of so-called $p$-nuclei. The standard calculations in the
statistical model show a dramatic sensitivity to the chosen $\alpha$-nucleus
potential. The present study explains the reason for this dramatic sensitivity
which results from the tail of the imaginary $\alpha$-nucleus potential in the
underlying optical model calculation of the total reaction cross section. As
an alternative to the optical model, a simple barrier transmission model is
suggested. It is shown that this simple model in combination with a
well-chosen $\alpha$-nucleus potential is able to predict total
$\alpha$-induced reaction cross sections for a wide range of heavy target
nuclei above $A \gtrsim 150$ with uncertainties below a factor of two. The new
predictions from the simple model do not require any adjustment of parameters
to experimental reaction cross sections whereas in previous statistical model
calculations all predictions remained very uncertain because the parameters of
the $\alpha$-nucleus potential had to be adjusted to experimental data. The
new model allows to predict the reaction rate of the astrophysically important
$^{176}$W($\alpha$,$\gamma$)$^{180}$Os reaction with reduced uncertainties,
leading to a significantly lower reaction rate at low temperatures. The new
approach could also be validated for a broad range of target nuclei from $A
\approx 60$ up to $A \gtrsim 200$.
\end{abstract}

\maketitle

%
{\it{Introduction.}}
The astrophysical \gpro\ is mainly responsible for the nucleosynthesis of
so-called \pnucs ; these are a group of heavy neutron-deficient nuclei with
very low abundances which are bypassed in the otherwise dominating neutron
capture processes \cite{Arnould_PhysRep2003}. The \gpro\ operates in an
explosive astrophysical environment at high temperatures of $2-3$ Giga-Kelvin
($T_9 = 2-3$).  Both, supernovae of type II
\cite{Woosley_APJS1978,Rauscher_APJ2002,Rauscher_MNRAS2016} and of type Ia
\cite{Travaglio_APJ2011,Travaglio_APJ2015,Nishimura_MNRAS2017} have been
suggested. Up to now, a final conclusion on the astrophysical site(s) of the
\gpro\ could not be reached. The combined uncertainties from the stellar
models and from the underlying nuclear reaction rates still prevent to
reproduce the abundances of all
\pnucs\ \cite{Rauscher_RPP2013,Pignatari_IJMPE2016}.

Nucleosynthesis in the \gpro\ proceeds via a series of photon-induced
reactions of \rgn , \rgp , and \rga\ type. In particular, most relevant for
the final abundances of the $p$-nuclei are the branching points between
\rgn\ and \rga\ which are typically located several mass units ``west'' of the
valley of stability for heavy $p$-nuclei and closer to stability for lighter
$p$-nuclei in the $A \approx 100$ mass region
\cite{Rapp_APJ2006,Rauscher_APJ2002,Rauscher_MNRAS2016,Nishimura_MNRAS2017,Rauscher_RPP2013,Pignatari_IJMPE2016}.
The astrophysical rates of photon-induced reactions are calculated
from the inverse capture reactions using detailed balance. It is generally
accepted that nucleon capture rates can be predicted with an uncertainty of
about a factor of two, whereas \al\ capture rates are more uncertain by at
least one order of magnitude (see e.g.\ the variation of rates in the
sensitivity study \cite{Rauscher_MNRAS2016}).

Astrophysically relevant energies, the so-called Gamow window, are of the
order of 10 MeV for heavy nuclei and temperatures of $T_9 \approx 2 - 3$. At
these sub-Coulomb energies the prediction of \al -induced reaction cross
sections is complicated because the usual statistical model calculations show
a wide range of predicted cross sections spanning over at least one order of
magnitude. This huge uncertainty results from the choice of the \al -nucleus
optical model potential (AOMP) in the statistical model (SM). For completeness
it has to be mentioned that the SM calculations are based on the total cross
section \stot\ which is calculated in the optical model (OM) by solving the
Schr\"odinger equation with a reasonable complex AOMP. Here \stot\ is given
by:
\begin{equation}
  \sigma_{\rm{reac}}(E) = \sum_L \sigma_L = {\frac{\pi \hbar^2}{2\mu E}}
  \sum_L (2L + 1) \left[ 1 - \eta^2_L(E) \right]
  \label{eq:stotOM}
\end{equation}
with the reduced mass $\mu$ and the (real) reflexion coefficient $\eta_L$ for
the $L$-th partial wave. Note that all $\eta_L = 1$ and thus \stot\ $ = 0$ in
the OM for any purely real AOMP without imaginary part.

The present study is organized as follows. In a first step, the main origin of
the huge uncertainties of \rag\ cross sections in the Gamow window is
identified for the first time. It will be shown that the imaginary part of the
AOMP at large radii (far outside the colliding nuclei) plays an essential
role. In a second step, a simple barrier transmission model will be suggested
which avoids the complications with the imaginary part of the AOMP. Next, this
simple model is combined with a carefully chosen AOMP, and total \al\ $+$
nucleus reaction cross sections are calculated at low energies, thus enabling
the prediction of \al -induced reaction rates for heavy target nuclei with
significantly reduced uncertainties. As a first example, \al -induced
reactions for \auvii\ were chosen because a recent experiment has provided
high-precision data down to energies close to the Gamow window
\cite{Szucs_PRC2019_au197aX}. Then, predictions of \al
-induced cross sections for several heavy targets (above $A = 150$) are
compared to experimental data from literature. Finally, a new prediction is
given for the reaction rate of the \wvi \rag \osnull\ reaction which governs
the production of the \pnuc\ \wnull\ \cite{Rauscher_MNRAS2016}; no
experimental data are available for the unstable target nucleus \wvi . The new
approach is also valid for target nuclei between $A \approx 60$ and $A \gtrsim
200$ (see Supplement \cite{Supplement}).

The present study uses spherical symmetry. The role of deformation for the
tunneling of \al\ particles was mainly investigated in \al -decay studies
(e.g.,
\cite{Delion_JPG2018_alphadecay,Xu_PRC2006_alphadecay,Delion_PRC2004_alphadecay}). Additional
information on the relevance of deformation is given in the Supplement
\cite{Supplement}.

{\it{Identification of the source of uncertainties.}}
Much work has been devoted to the determination of global AOMPs at low
energies in the recent years. Starting with the pioneering work of Somorjai
{\it et al.}\ \cite{Somorjai_AaA1998_sm144ag} on the $^{144}$Sm\rag $^{148}$Gd
reaction, it was noticed that the available AOMPs overestimate the
experimental data in particular towards low energies. This holds for the
widely used simple AOMP by McFadden and Satchler (MCF)
\cite{McFadden_NPA1966_aomp} and the early AOMP by Watanabe
\cite{Watanabe_NPA1958_omp} which was the default choice in previous versions
of the widely used SM code TALYS \cite{TALYS-V19}. Nowadays, TALYS offers a
broader choice of AOMPs, including the AOMPs
by Avrigeanu {\it et al.}\ \cite{Avrigeanu_PRC2014_aomp}
(present default choice in TALYS) and
by Demetriou {\it et al.}\ \cite{Demetriou_NPA2002_aomp}.
Furthermore, we modified TALYS to implement
the recent ATOMKI-V1 potential \cite{Mohr_ADNDT2013_atomki-v1}. It was found
that the total reaction cross section of \al\ $+$ \auvii\ from the different
AOMPs varies by less than a factor of two at higher energies of 25 MeV above
the Coulomb barrier; however, around 10 MeV, i.e., in the center of the Gamow
window at $T_9 \approx 2.5$, the predicted cross sections vary dramatically by
about four orders of magnitude \cite{Szucs_preprint_NPA9}. Obviously, the
reason for the wide range of predictions should be understood.

For explanation, we start with the simple 4-parameter MCF potential which uses
a standard Woods-Saxon (WS) parametrization with the real and imaginary depths
of $V_0$ = 185 MeV, $W_0 = 25$ MeV, radius $R = 1.4$ fm $\times A_T^{1/3} =
8.15$ fm, and diffuseness $a = 0.52$ fm. Three changes are applied to the MCF
potential; these changes show that the tail of the imaginary part of the
potential far outside the colliding nuclei has the dominating influence on the
calculated low-energy cross sections.

\begin{figure}[htb]
\includegraphics[width=\columnwidth,clip=]{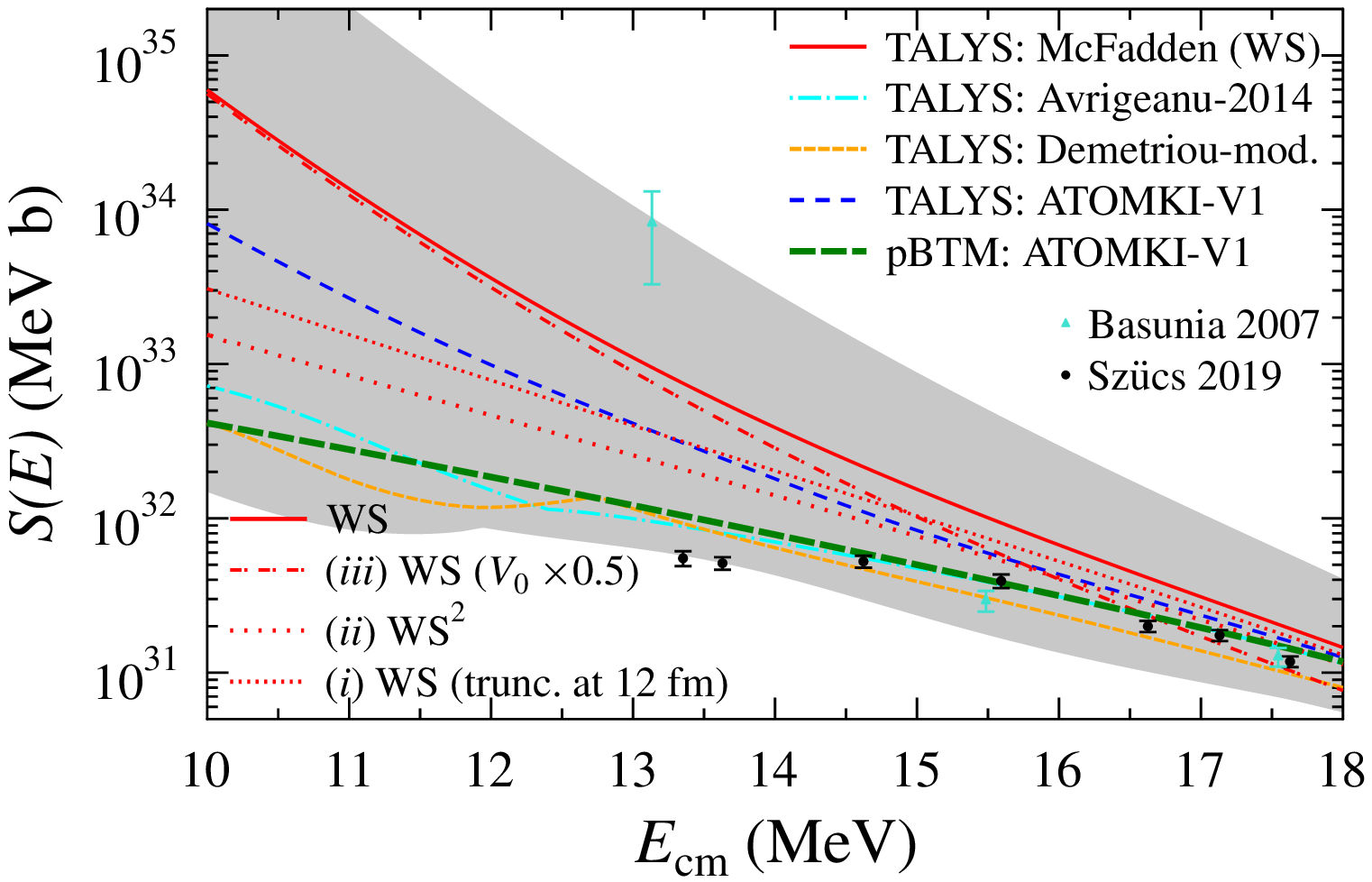}
\caption{
\label{fig:au_sigtot}
(Color online)
Total reaction cross section \stot\ for \al\ $+$ \auvii\ for different
potentials (shown as astrophysical \sfact ), compared to experimental data
which are taken from the sum of \rag , \ran , and \rann\ in
\cite{Basunia_PRC2007,Szucs_PRC2019_au197aX}. (The comparison of calculated
total reaction cross sections \stot\ to the sum of partial experimental cross
sections avoids any complications from other ingredients of the SM
calculations like the \g -strength function or the level density.)
Modifications ($i$) and ($ii$) of the imaginary MCF WS potential reduce
\stot\ at low energies dramatically (dotted red lines); the reduction of the
real MCF potential ($iii$) has minor influence at low energies (dash-dotted
red). The TALYS and pBTM calculations are discussed in the text. The shaded
area represents the wide range of TALYS predictions.
}
\end{figure}

($i$) We truncate the imaginary part of the MCF WS potential at $r = 12$ fm
(with the tiny $W(r) \approx 0.015$ MeV or $W(r)/W_0 < 10^{-3}$!). This
truncation has minor influence at higher energies but reduces \stot\ at low
energies by one order of magnitude (dotted line in Fig.~\ref{fig:au_sigtot}).

($ii$) We change the parameterization of the imaginary part to a squared
Woods-Saxon (WS$^2$) and re-adjust the parameters of the WS$^2$ potential such
that the imaginary potential is practically identical to the initial MCF WS
potential up to about 10 fm, but significantly weaker at larger radii; for the
WS$^2$ potential we find $W_0 = 25.07$ MeV, $R_0 = 1.499$ fm, $a = 0.623$
fm. This results in a similar energy dependence for \stot\ as in the previous
case ($i$), see wide-dotted line in Fig.~\ref{fig:au_sigtot}.

($iii$) We reduce the real part of the MCF WS potential by a significant
factor of two. This reduction increases the effective barrier, and thus -- as
expected -- \stot\ is reduced at higher energies (dash-dotted line in
Fig.~\ref{fig:au_sigtot}). However, around 10 MeV \stot\ does not change
because \stot\ results from the tail of the imaginary potential; i.e.,
absorption (in the OM calculation) occurs at large radii far outside the
colliding nuclei, before the incoming \al\ has tunneled through the
barrier. Consequently, the height of the barrier is practically not relevant
at the lowest energies in Fig.~\ref{fig:au_sigtot}.

The extreme sensitivity to the tail of the imaginary potential at large radii
is the simple explanation for the huge range of predictions of low-energy
\stot\ for \al -induced reactions from different AOMPs (shown as gray-shaded
area in Fig.~\ref{fig:au_sigtot}). It has to be noted that the shape of the
imaginary potential is usually fixed by the analysis of elastic scattering at
energies around and above the Coulomb barrier. These experiments, however, do
not constrain the tail of the imaginary part. For example, at an energy of 25
MeV, the WS and WS$^2$ potentials in cases ($ii$) and ($iii$) provide
\stot\ within 2\%, and the deviation in the calculated angular distributions
never exceeds 9\% in the full angular range. In practice, the tail of the
imaginary potential results -- more or less by accident -- from the chosen
parametrization in the fitting of the elastic angular distribution where the
parameters of the imaginary part are mainly sensitive to the nuclear surface
region but not to the far exterior.

%
{\it{An alternative approach.}}
From the above discussion it is obvious that any calculation of the total
reaction cross section \stot\ in the OM at energies far below the Coulomb
barrier must have significant uncertainties. Here we present an alternative
approach which avoids the uncertainties from the unknown imaginary potential
at large radii. A similar approach -- extended by coupling to low-lying
excited states -- is widely used for heavy-ion fusion reactions, and there it
was found that WS potentials are inappropriate to describe data far below the
Coulomb barrier
\cite{Back_RMP2014_fusion,Hagino_PTP2012_fusion,Balantekin_RMP1998_fusion}.

The suggested model is based on the calculation of transmission through the
Coulomb barrier in a purely real nuclear potential; it will be called ``pure
barrier transmission model'' (pBTM) in the following. By definition, this
model assumes absorption of an incoming \al -particle, as soon as the \al\ has
tunneled through the barrier from the exterior to the interior. This
assumption is reasonable because the small tunneling probability of the \al
-particle prevents the \al\ from tunneling back to the exterior; it is much
more likely that the formed compound nucleus decays by \g -ray or neutron
emission. The total cross section in the pBTM is given by
\begin{equation}
  \sigma_{\rm{reac}}(E) = \sum_L \sigma_L = {\frac{\pi \hbar^2}{2\mu E}}
  \sum_L (2L + 1) T_L(E)
  \label{eq:stotBTM}
\end{equation}
with the barrier transmission $T_L$; for
comparison, see also Eq.~(\ref{eq:stotOM}) for \stot\ in the OM.

Technically, the calculations in the pBTM were performed using the code CCFULL
\cite{Hagino_CPC1999_ccfull}. Minor modifications to the code had to be made
to use numerical external potentials. For further technical details of the
pBTM and the calculations, see also \cite{Mohr_IJMPE2019_40ca} and the
Supplement \cite{Supplement}.

The real part of the full ATOMKI-V1 potential is energy-independent whereas
the imaginary part increases with energy around the barrier. The resulting
coupling between the imaginary and real parts is governed by the so-called
dispersion relation
\cite{Nagarajan_PRL1985_disp,Nagarajan_PLB1986_disp,Mahaux_NPA1986_disp,Mahaux_NPA1986_disp2}.
It is found that the additional consideration of the dispersion relation has
only minor influence of less than 30\% on the total cross section \stot\ for
all energies under study because the parameters of the chosen ATOMKI-V1
potential were adjusted to elastic scattering data at energies around the
Coulomb barrier. A study of dispersion relations is provided in the Supplement
\cite{Supplement}.

The results for \auvii\ $+$ \al\ from the simple pBTM are compared to SM
calculations using different AOMPs from TALYS in
Fig.~\ref{fig:au_sigtot}. Obviously, the SM calculations using the MCF
and the ATOMKI-V1 potentials overestimate the experimental low-energy cross
sections. Lower cross sections result from the Avrigeanu AOMP
\cite{Avrigeanu_PRC2014_aomp} and from a Demetriou AOMP
\cite{Demetriou_NPA2002_aomp}; the latter has been scaled by a factor of 1.2 (as
suggested in \cite{Scholz_PLB2016_aomp}). Interestingly, the simple pBTM
model in combination with the real part of the ATOMKI-V1 potential leads to
cross sections which are close to the experimental data and also close to the
many-parameter potentials by Avrigeanu {\it et
  al.}\ \cite{Avrigeanu_PRC2014_aomp} and Demetriou {\it et
  al.}\ \cite{Demetriou_NPA2002_aomp,Scholz_PLB2016_aomp}. 

\begin{figure}[htb]
\includegraphics[width=0.925\columnwidth,clip=]{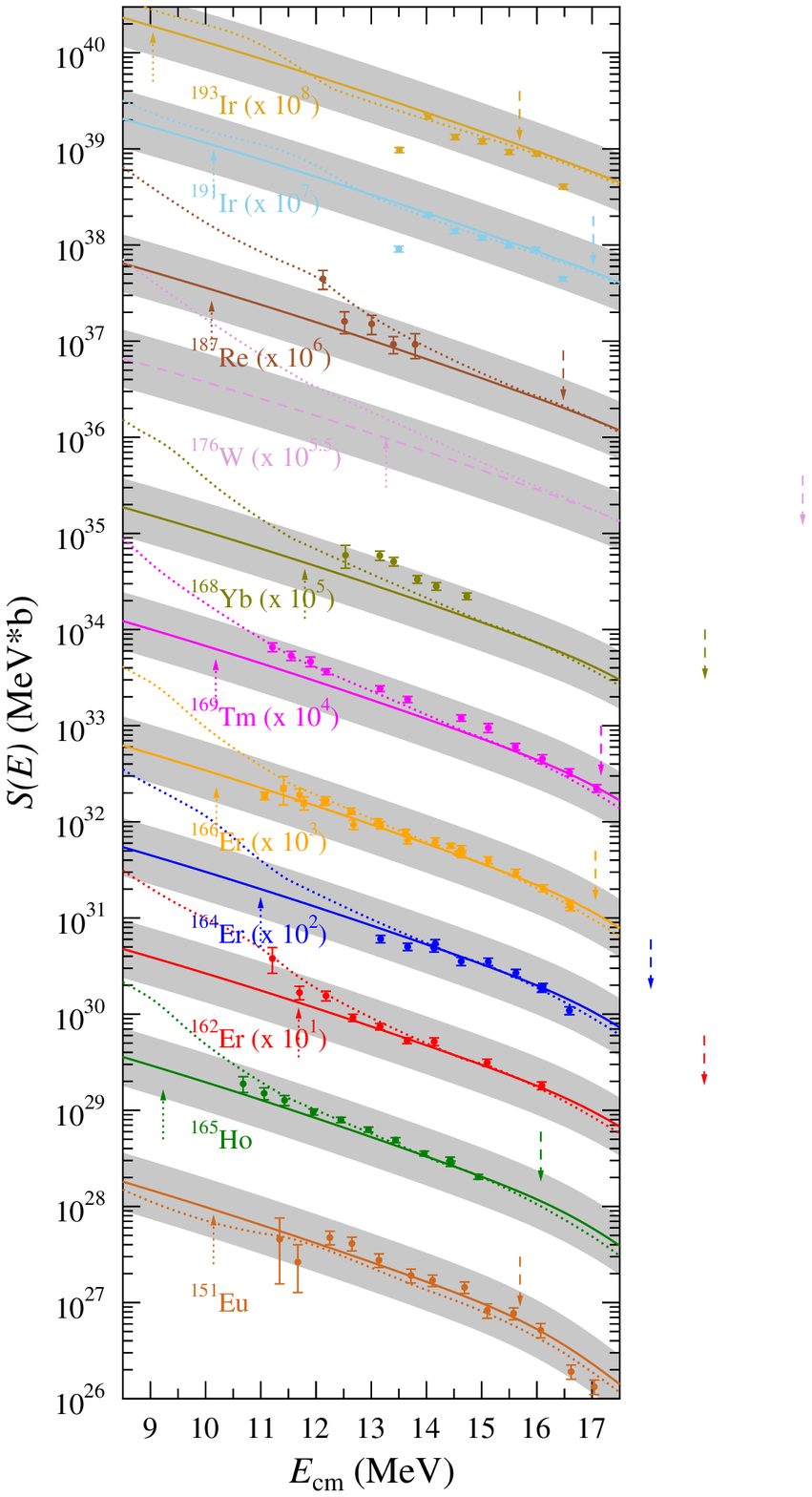}
\caption{
\label{fig:above150}
(Color online)
Total reaction cross section \stot\ (given as astrophysical \sfact ) for \al
-induced reactions above $A = 150$. The pBTM predicts practically
all experimental data \cite{Gyurky_JPG2010_151eu,Glorius_PRC2014_165ho_166er,Kiss_PLB2014_162er,Kiss_JPG2015_er,Kiss_PLB2011_169tm,Netterdon_NPA2013_168yb,Scholz_PRC2014_187re,szucs_PLB2018_ir}
within a factor of two (grey shaded). The dotted lines show the
results from the many-parameter AOMP of \cite{Avrigeanu_PRC2014_aomp}. The
arrows indicate the \ran\ and \rann\ thresholds.
}
\end{figure}

Encouraged by this successful application of the pBTM for \auvii\ $+$ \al , we
have calculated \stot\ for a series of \al -induced reactions of heavy target
nuclei above $A = 150$. Fig.~\ref{fig:above150} shows that the predictions
from the pBTM are in excellent agreement with recent experimental data
\cite{Gyurky_JPG2010_151eu,Glorius_PRC2014_165ho_166er,Kiss_PLB2014_162er,Kiss_JPG2015_er,Kiss_PLB2011_169tm,Netterdon_NPA2013_168yb,Scholz_PRC2014_187re,szucs_PLB2018_ir}. Typical deviations are less than a factor of two which is marked as
grey-shaded uncertainty band in Fig.~\ref{fig:above150}. No parameter
adjustment to reaction cross sections is necessary for the present
calculations because the real part of the ATOMKI-V1 potential is completely
constrained from elastic scattering and an imaginary part is not required in
the simple pBTM. Technical details on the calculation of the double-folding
potential ATOMKI-V1 and the chosen density distributions are provided in the
Supplement \cite{Supplement}.

We benchmark the calculations in the simple pBTM with the results from the
AOMP by Avrigeanu {\it et al.}\ \cite{Avrigeanu_PRC2014_aomp} (shown as dotted
lines in Fig.~\ref{fig:above150}). This many-parameter AOMP ($\gg 10$
parameters, see Table II of \cite{Avrigeanu_PRC2014_aomp}) has been adjusted
to most of the experimental data shown in
Fig.~\ref{fig:above150}. Interestingly, a minor enhancement of the imaginary
potential was introduced in \cite{Avrigeanu_PRC2014_aomp} for $152 \le A
\le 190$, leading to a significantly increased low-energy \sfact\ which is not
present for $^{151}$Eu and $^{191,193}$Ir. Contrary to the many-parameter
approach of \cite{Avrigeanu_PRC2014_aomp}, no adjustment of parameters is
required in the present pBTM; nevertheless, the deviation from the
experimental data is typically less than a factor of two.

\begin{figure}[htb]
\includegraphics[width=\columnwidth,clip=]{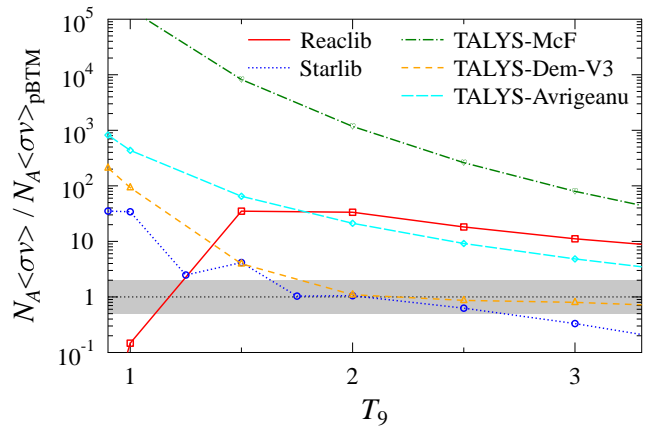}
\caption{
\label{fig:w176rate}
(Color online)
Reaction rate \Nsv\ of the \wvi \rag \osnull\ reaction, normalized to the
present calculation in the pBTM. The grey shaded area indicates the
uncertainty of a factor of two (see also Fig.~\ref{fig:above150}). Further
discussion see text and the Supplement \cite{Supplement}.
}
\end{figure}

Again encouraged by the successful application of the simple pBTM model to $A
> 150$ nuclei, we finally predict the reaction rate of the \wvi \rag
\osnull\ reaction which is essential for the nucleosynthesis of the
\pnuc\ \wnull\ \cite{Rauscher_MNRAS2016}. Because of the highly negative
$Q$-value of the \ran\ channel, the total cross section is approximately
identical to the \rag\ cross section in the astrophysically relevant energy
range. Thus, the total cross section \stot\ from the pBTM can be directly
used for the calculation of the reaction rate \Nsv\ of the
\rag\ reaction. The result from the pBTM is compared to other predictions
\cite{Cyburt_APJS2010_reaclib,Rauscher_ADNDT2000_rates,Sallaska_APJS2013_starlib}
in Fig.~\ref{fig:w176rate}. The rates from literature cover several orders of
magnitude, even exceeding the the range of variations in the
sensitivity study \cite{Rauscher_MNRAS2016}, whereas the present approach
should be valid within a factor of two. For further details, see the
Supplement \cite{Supplement}.

%
{\it{Summary and conclusions.}}
The present work has identified the reason for the huge variations of \al
-induced reaction cross sections at low energies in the statistical model
which results from the tail of the imaginary part of the \al -nucleus
potential. As an alternative to the statistical model, a simple barrier
transmission model is suggested where the total reaction cross section is
calculated from the transmission through the Coulomb barrier in a real
potential. The combination of this simple barrier transmission model with the
real part of the ATOMKI-V1 potential leads to predictions of total \al
-induced cross sections which agree with the experimental data within less
than a factor of two for a wide range of heavy target nuclei above $A > 150$.
Contrary to previous approaches, the present calculations do not require any
adjustment of parameters and thus predict low-energy cross sections from a
simple, but physically sound model.

The new approach is used to predict the reaction rate of the astrophysically
important \wvi \rag \osnull\ reaction which has strong impact on the abundance
of the \pnuc\ \wnull\ \cite{Rauscher_MNRAS2016}. According to the small
deviations from the experimental data for all targets under study, we claim an
uncertainty of less than a factor of two for this rate whereas previous
predictions of \Nsv\ are higher than the present result and vary by orders of
magnitude.

The present study focuses on heavy target nuclei with masses above $A >
150$ where the predictions from different \al -nucleus potentials vary over
orders of magnitude. For lighter targets, the predictions of \al -induced
cross sections from different potentials do not vary as dramatic, and it was
found that also the simple barrier transmission model reproduces experimental
data very well. The recently measured $^{100}$Mo\ran $^{103}$Ru data
were predicted with similar uncertainties as in the $A \gtrsim 150$ mass range
\cite{Szegedi_preprint_NPA9}, and data for \zniv\ + \al\ and \niviii\ +
\al\ were also reproduced. Further information on the applicability of the
barrier transmission model in a wide mass range and for nuclei beyond the
valley of stability is provided in the Supplement \cite{Supplement}.
In conclusion, the present approach is valid for masses above $A \ge 58$, and
thus a reliable prediction of \al -induced reaction cross sections comes
within reach for the whole nucleosynthesis network of the \gpro . Furthermore,
the present approach is also able to provide improved reaction rates for
\ran\ reactions in the weak
\rpro\ \cite{Pereira_PRC2016_a-n,Mohr_PRC2016_a-n,Bliss_JPG2017,Bliss_PRC2020}. 

The Supplement \cite{Supplement} provides the following additional references:
\cite{deVries_ADNDT1987,Feshbach_annPhys1958_omp,Abele_PRC1993_16o,Gomes_PRC2005_reduced,Quinn_PRC2014_58ni,McGowan_PR1964_58ni_59co_a-X,Vlieks_NPA1974_a-X,Cummings_PR1959,Stelson_PR1964_a-n,Mohr_EPJA2015_A20-50,Gyurky_PRC2012_64zn,Ornelas_PRC2016_64zn,Mohr_PRC2017_64zn,Trache_EPJWebConf2020_ni,Barnett_PRC1974_a-X,Pereira2019_priv,Avila2019_priv,Rauscher_PRC2010_gamow}.

{\it{Acknowledgments}}
We thank E.\ Somorjai, T.\ Rauscher, and D.\ Galaviz for countless encouraging
discussions on \al -induced reaction cross sections over more than two decades.
This work was supported by NKFIH (Gr. No. K120666, NN128072) and by the New
National Excellence Program of the Ministry for Innovation and Technology
(\'UNKP-19-4-DE-65).
G.\,G.~Kiss acknowledges support from the J\'anos Bolyai research
fellowship of the Hungarian Academy of Sciences.

\bibliography{pm}

\end{document}